\documentstyle[epsfig,mprocl]{article}

\def\Journal#1#2#3#4{{#1} {\bf #2}, #3 (#4)}
\def\NPB{{\em Nucl. Phys.} B}

\def\PRL{\em Phys. Rev. Lett.}
\def\PRD{{\em Phys. Rev.} D}

\def\CQG{{\em Class. and Quantum Grav.}} 
\def\JMP{{\em Jour. Math. Phys.}} 
\def\be{\begin{equation}}
\def\ee{\end{equation}}
\def\bea{\begin{eqnarray}}
\def\eea{\end{eqnarray}}
\begin{document}

.

\vspace{-0.9in} 
\vbox{
%%%\baselineskip=15 pt 
%%%\epsffile{header.ps} 
%%%  
\hfill\vbox{ \hbox{CGPG-97/11-1}
             \hbox{University of Parma UPRF 97-13}
             \hbox{xxx-archive gr-qc/9711021}
}}
\vspace{.2in} 

\title{THE EQUIVALENCE BETWEEN THE CONNECTION AND
       THE LOOP REPRESENTATION OF QUANTUM GRAVITY}

\author{R. De Pietri}

\address{Dipartimento di Fisica, Universit\`a degli Studi di Parma, and 
INFN, \\ Gruppo Collegato di Parma, viale delle scienze, 43100 Parma, Italy.\\
and \\
Center for Gravitational Physics an Geometry, 
The Pennsylvania State University, \\
104 Davey Laboratory, University Park, PA 16802-6300, USA}

%%%%%%%%%%%%%%%%%%%%%%%%%%%%%%%%%%%%%%%%%%%%%%%%%%%%%%%%%%%%%%
% you may repeat \author \address as often as necessary      %
%%%%%%%%%%%%%%%%%%%%%%%%%%%%%%%%%%%%%%%%%%%%%%%%%%%%%%%%%%%%%%

\maketitle\abstracts{
The recent developments of the ``connection'' and ``loop'' 
representations have given the possibility to show that the two 
representations are equivalent and that it is possible to transform 
any result from one representation into the other.  
The glue between the two representations is the loop transform.
Its use, combined with  Penrose's binor
calculus, gives the possibility of establishing the exact 
correspondence between  operators and states in the 
connection representation and those in the loop
representation. \\
The main ingredients in the proof of the equivalence are: the concept of
embedded spin network, the Penrose graphical method of
$SU(2)$ calculus, and the existence of a generalized measure
on the space of connections.}
  
%%%%%%%%%%%%%%%%%%%%%%%%%%%%%%%%%%%%%%%%%%%%%%%%%%%
%%    SECTION
%%%%%%%%%%%%%%%%%%%%%%%%%%%%%%%%%%%%%%%%%%%%%%%%%%%
%%%\section{Introduction}

The {\it loop}\cite{LOOP} and the {\it connection}\cite{AI}
representations  approch to the canonical quantizion of GR 
(Einstein's general relativity) have the goal of constructing
a quantum theory, based on connection $A_a^i(x)$ modulo gauge transformations 
(in the case of GR the  Ashtekar-Sen $SU(2)$ connection),
such that the Wilson's loop functionals (the trace of the Holonomy 
along a closed loop $\alpha$)
\be
{\cal T}_\alpha[A] = {\rm Tr}{\cal P} 
   \exp\big[\int_\alpha A_a(x) dx^a  \big]
\ee
become well defined quantum operators. 
In the  {\it loop-representation} approch  the quatization is achieved 
realizing the {\it quantum operator}  that correspond to the ${\cal T}$ 
observables on the vector space ${\cal V}_{loop}$ of 
all the loops modulo the Mandelstam 
relations.\footnote{It is important to note that in this approch 
the essential problem of the definition of the scalar product
and indeed of the the Hilbert Space structure was postpone
(see \cite{DePietri96} for the solution of this problem on the original
phylosophy of the {\it loop-representation}).}
In the {\it connection-representation} approch, in contrast, the 
first step was the construction of the Hilbert space structure
in which the ${\cal T}_\alpha[A]$ operators are realized as
multiplications: the Hilbert space 
${\cal H}=L^2[\overline{{\cal A/G}},d\mu]$ 
of the square integrable  function with respect to the Gel'fand 
spectral measure associated to the $C^\star$ algebra of 
the ${\cal T}_\alpha[A]$'s. These two approaches are connected by 
the so-called loop trasformation. To any state 
$\psi_C \in {\cal H}$, it is assocated a state 
$\psi_L \in {\cal V}_{loop}$ as:
\be \label{LT}
  \psi_L(\alpha) = <\alpha|\psi > 
  = \int_{\overline{\cal A/G}} d\mu(A)  <\alpha | A> <A | \psi > 
  = \int_{\overline{\cal A/G}} d\mu(A) \overline{T_\alpha[A]} \psi_C(A)
~~.
\label{Ltransf}
\ee
The problem of proving the equivalence of the two representations
is equivalent to the problem of showing the explicit action 
of this transformation. The different mathematical framework 
of the two representations was the only reason behind the 
difficulty.

This problem  has a straightforward solution using Penrose's 
graphical binor calculus for the $SU(2)$-tensors in the 
connection representation\cite{DePietri96a}. Using this method, 
it is immediate to show that the loop  transform (\ref{LT}) maps 
the spin-network basis of ${\cal V}_{loop}$\cite{RS} into the
spin-network basis of ${\cal H}=L^2[\overline{{\cal A/G}},d\mu]$
\cite{CONNECTION}. We refer the interest reader to 
\cite{DePietri96a} for a detailed  account of the proof and 
for the relevant bibliography.

%%%This note is diveded in 4 schematic paragraph: 
%%%\P 1 characterizes the space ${\cal H}$; \P 2 
%%%gives an account of Penrose's binor calculus; 
%%%\P 3 defines the binor representation of the space
%%%${\cal H}$ and final \P 4 discusses the equivalence 
%%%of the two spin-nettwork bases.
%%%%\section{The space ${\cal H}$ and the binor graphical representation.}
%%%%

The basic idea behind Penrose's binor calculs is to represent any
$SU(2)$ tensor (i.e., tensor expression with indices $A,B,\ldots=1,2$) 
in terms of the following graphical elements in the plane:
\begin{equation}
\delta_C^{~A} = 
   \begin{array}{c}\setlength{\unitlength}{1 pt}
   \begin{picture}(10,20)
            \put(5,5){\line(0,1){10}}
            \put(5,5){\circle*{3}}\put(5,15){\circle*{3}}
            \put(6,0){${\scriptstyle C}$}\put(6,14){${\scriptstyle A}$}
   \end{picture}\end{array} 
~~~~
{\rm i} \epsilon_{AC} = 
   \begin{array}{c}\setlength{\unitlength}{1 pt}
   \begin{picture}(20,15)
               \put(5,5){\circle*{3}}\put(15,5){\circle*{3}}
               \put(10,5){\oval(10,20)[t]}
               \put(3,0){${\scriptstyle A}$}\put(12,0){${\scriptstyle C}$}
   \end{picture}\end{array}
~~~~
{\rm i} \epsilon^{AC} = 
   \begin{array}{c}\setlength{\unitlength}{1 pt}
   \begin{picture}(20,15)
         \put(10,10){\oval(10,20)[b]}
         \put(5,10){\circle*{3}}\put(15,10){\circle*{3}}
         \put(3,11){${\scriptstyle A}$}\put(12,11){${\scriptstyle C}$}
   \end{picture}\end{array} 
~~~~
X_{AB}^{C} =
   \begin{array}{c}\setlength{\unitlength}{1 pt}
   \begin{picture}(20,25)
      \put(0,7){\framebox(20,10){${\scriptstyle X}$}}
      \put(6,0){${\scriptstyle A}$}\put(5,2){\line(0,1){5}}
      \put(5,2){\circle*{3}}
      \put(16,0){${\scriptstyle B}$}\put(15,2){\line(0,1){5}}
      \put(15,2){\circle*{3}}
      \put(11,20){${\scriptstyle C}$}\put(10,17){\line(0,1){5}}
      \put(10,22){\circle*{3}}
   \end{picture}\end{array}
\label{eq:binCONV}
\end{equation}
and assigns to any crossing a minus sign, i.e:
$X^{AB}_{CD} = \delta_D^{~A}\delta_C^{~B} = - 
   \begin{array}{c}\setlength{\unitlength}{1 pt}
   \begin{picture}(24,10)
        \put(7,0){\line(1,1){10}}
        \put(7,0){\circle*{3}}
        \put(7,10){\circle*{3}}
        \put(0,0){${\scriptscriptstyle C}$}
        \put(0,8){${\scriptscriptstyle A}$}
        \put(17,0){\line(-1,1){10}}
        \put(17,0){\circle*{3}}
        \put(17,10){\circle*{3}}
        \put(19,0){${\scriptscriptstyle D}$}
        \put(19,8){${\scriptscriptstyle B}$}
   \end{picture}\end{array}
$. Using this rule it is possible to represent any $SU(2)$
($SL(2,C)$) tensor expression in a graphical way. 
In particular we have the following graphical representation 
for {(\it i}) the irreducible representation $\pi_i(n_i)$
\footnote{Labeled by an integer $n$, its color, that is twice the 
spin: $n = 2 j_n$}, and of the unique 3-valent contractor
\begin{eqnarray}
&&  \pi_i(n_i)   
=  \begin{array}{c}\setlength{\unitlength}{1 pt}
   \begin{picture}(15,30)
    \put(7,22){$\scriptstyle n_i$}
    \put(10,12){$\scriptstyle {e_i}$}
    \put(2,5){\rule{6pt}{15pt}}
    \put(5,0){\line(0,1){5}}
    \put(5,20){\line(0,1){5}}
   \end{picture}\end{array}
=  \begin{array}{c}\setlength{\unitlength}{1 pt}
   \begin{picture}(40,40)
    \put(22,2){$\scriptstyle n_i$}
    \put(0,10){\framebox(40,2){}} \put(20,0){\line(0,1){10}}
    \put(5,12){\line(0,1){3}}\put(5,25){\line(0,1){3}}
    \put(-1,15){\framebox(12,10){$\scriptstyle g_{e_i}$}}
    \put(12,18){$\cdots$}
    \put(35,12){\line(0,1){3}}\put(35,25){\line(0,1){3}}
    \put(29,15){\framebox(12,10){$\scriptstyle g_{e_i}$}}
    \put(0,28){\framebox(40,2){}} \put(20,30){\line(0,1){10}}
    \put(22,34){$\scriptstyle n_i$}
   \end{picture}\end{array}
~,~~~~ 
\Pi^{(e)}_n P_n = \frac{1}{n!} \sum_p
(-1)^{|p|}\ P^{(p)}_{p}
=  \begin{array}{c}\setlength{\unitlength}{1 pt}
   \begin{picture}(20,25)
     \put(10, 0){\line(0,1){10}}
     \put(0,10){\framebox(20,5){}}
     \put(10,15){\line(0,1){10}}\put(12,17){n}
   \end{picture}\end{array}
~,
\nonumber
\\ &&
%%%%%%%%%%%%%%%%%%%%%%%%%%%%%%%%%%%%%%%%%%%%%%%%%%%%%%%%%%%%%%%%
   \begin{array}{c}\setlength{\unitlength}{1 pt}
   \begin{picture}(30,40)
       \put(15,15){\line(-1, 1){10}} \put( 4,27){$a$}
       \put(15,15){\line( 1, 1){10}} \put(22,27){$b$}
       \put(15, 5){\line(0,1){10}}   \put(17,1){$c$}
       \put(15,15){\circle*{3}}
   \end{picture}\end{array}
\stackrel{def}{=}
   \begin{array}{c}\setlength{\unitlength}{1 pt}
   \begin{picture}(70,40) 
       \put(20,30){\line( 1, 0){30}} \put(28,32){$m$}
       \put(35,15){\line(-1, 1){15}} \put(20,17){$p$}
       \put(35,15){\line( 1, 1){15}} \put(44,17){$n$}
       \put(35, 1){\line(0,1){10}}   \put(42,1){$c$}
       \put(27,11){\framebox(16,4){}}
       \put(16,22){\framebox(4,16){}} 
       \put(6 ,30){\line(1,0){10}}\put(0,30){a}
       \put(50,22){\framebox(4,16){}}
       \put(54,30){\line(1,0){10}}\put(65,30){b}
   \end{picture}\end{array}
\; ~,\qquad
\left\{\begin{array}{rcl} 
m &=& (a+b-c)/2 \\ n&=&(b+c-a)/2 \\ p&=&(c+a-b)/2
\end{array}\right.
~~~.
\label{eq:6}
\end{eqnarray}

Now, the space ${\cal H}=L^2[\overline{{\cal A/G}},d\mu]$
and its spin-network basis are defined as follows: 
({\bf i}) the quantum configuration space 
   $\overline{{\cal A}/{\cal G}}$ is
   taken to be the Gel'fand spectrum generated by the Wilson
   loop functionals;
({\bf ii}) the space $\overline{{\cal A}/{\cal G}}$ could
   be characterized as the projective limit of the 
   finite dimensional spaces $\overline{{\cal A}/{\cal G}}_\gamma$
   of the cylindrical functions associated to piecewise
   analytical graphs $\gamma$ and in this space a fiducial 
   measure $d\mu_0(A)$ is  naturally defined as the 
    $\sigma$-additive extension of the  family of 
   products of Haar measures 
   $d\mu_{0,\gamma}(A)=d\mu_H(g_{e_1})\ldots d\mu_H(g_{e_n})$
   in the spaces $\overline{{\cal A}/{\cal G}}_\gamma$; 
   A function $f_\gamma$ ($f_\gamma \in \overline{{\cal A}/{\cal G}}_\gamma$)
   is said to be cylindrical with respect to a graph $\gamma$ if it is a gauge
   invariant function of the finite set of arguments
   $(g_{e_1}(A),\ldots,g_{e_n}(A))$ where the $g_{e_i} =
   {\cal P}\exp(-\int_{e_i} A)$ are the holonomies of $A$ along the edges
   $e_i$ of the graph $\gamma$. 
({\bf iii}) a natural basis in the space $\overline{{\cal A}/{\cal G}}$
is given by the spin-network cylindrical functions. 
They express the fact that any function cylindrical with respect to 
a graph $\gamma$ can be decomposed in terms of irreducible representations,
i.e. 
\[ 
%%\begin{equation}
f_\gamma(g_{e_1},\ldots,g_{e_n}) = \!\sum_{\vec{\pi},\vec{c}} \!
   f(\gamma,\vec{\pi},\vec{c}) ~~{\cal T}_{\gamma,\vec{\pi},\vec{c}}[A]
,~~{\cal T}_{\gamma,\vec{\pi},\vec{c}}[A] \stackrel{def}{=}
  \left( \otimes_{i=1}^{\#^{\rm edge}}\! \pi_i(g_{e_i}) \right) \cdot
  \left( \otimes_{j=1}^{\#^{\rm vetex}}\! c_j \right) 
\] 
%%\label{eq:defSPINnet} \end{equation}
where: {\bf (i)} $\vec{\pi}=(\pi_1,\ldots,\pi_N)$ denotes the 
labeling of the edges with irreducible 
representation $\pi_i$ of $G$; 
{\bf (ii)} $\vec{c}=(c_1,\ldots,c_M)$ a labeling  of
    the vertices with invariant contractors $c_j$ 
(the intertwining matrices $c_j$, in each of the  vertices $v_j$, 
represent the invarian coupling of the $n_j$ representations associated 
to the $n_j$ edges that start or end in $v_j$).

%%%%%%%%%%FIG 1%%%%%%%%%%%%%%%%%%%%%%%%%%%%%%%%%%%%%%%%%%%%%%%%%%%%%
\begin{figure}
\begin{center}
\mbox{\psfig{file=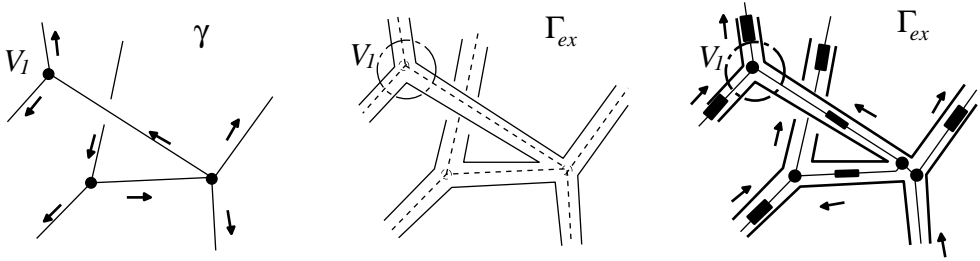}} 
\end{center}
\caption{The graph $\gamma$, a possible extended-planar 
projection $\Gamma_{ex}$ and the graphical representation
of a spin-network cilyndrical function}
\label{fig:graph}
\end{figure}
%%%%%%%%%%%%%%%%%%%%%%%%%%%%%%%%%%%%%%%%%%%%%%%%%%%%%%%%%%%%%%%%%%

At this point, it is immediate to define the graphical 
binor representation of the spin network 
${\cal T}_{\gamma,\vec{\pi},\vec{c}}[A]$. 
Refering to Fig.1, (1) cosider a planar projection
of the graph $\gamma$ and (2) its extended planar
projection $\Gamma_{ex}$; (3) insert in any extended-edge the  
graph-rep.\ of the irreduciple representation and in 
any extended-vertex the graph-rep.\ of the contractor in terms
of its decomposition in terms of 3-valent invariant tensor;
(4) represent index contraction as the joining of the corresponding 
lines.

%%%\section{The loop transform of the {\it loop} spin-network bases 
%%%         and the equivalence of the two representations}

Now, consider the definition of the spin network state in the 
loop representation given in \cite{DePietri96}.
We are left with the task of proving that the loop-transform 
of them is exactly a spin-network state of the connection 
representation. Refering to eq. (\ref{Ltransf}) we have to show
$<A,\alpha>_{Loop} = {T_{\gamma,\vec{\pi},\vec{c}}[A]}$.
Recalling that a spin network in the loop representation
(section V of \cite{DePietri96}) is exactly the drawing on 
$\Gamma_{ex}$ corresponting to the graphical binor-representation
of a spin-network basis of the connection representation 
(in the normalization discused in the previous section), the
assertion that the Loop-Transform of a spin-network of the loop
representation is a spin-network of the connection representation
follow in a straightforward way.

I thank Carlo Rovelli for a critical reading of the manuscript,
suggestions, help, encouragment and the uncountable number of discussion
we had about ``Quantum Gravity'' in the last years.

%%%%%%%%%%%%%%%%%%%%%%%%%%%%%%%%%%%%%%%%%%%%%%%%%%%
%%%%%%%%%%%%   references  %%%%%%%%%%%%%%%%%%%%%%%%
%


\section*{References}
\begin{thebibliography}{99}

\bibitem{AI} A Ashtekar and C J Isham
   \Journal{\CQG}{9}{1433}{1992}.

\bibitem{LOOP} 
  C Rovelli and L Smolin  \Journal{\PRL}{61}{1155}{1988} and 
  \Journal{\NPB}{331}{80}{1990}; 

\bibitem{CONNECTION} A Ashtekar, J Lewandowski, D Marolf, 
   J Mour\~ao and T. Thiemann \Journal{\JMP}{36}{6456}{1995}.

\bibitem{DePietri96a} R De~Pietri \Journal{\CQG}{14}{53}{1997}.

\bibitem{RS} C Rovelli and L Smolin
   \Journal{\PRD}{54}{5743}{1995}.

\bibitem{DePietri96} R De Pietri and C Rovelli 
   \Journal{\PRD}{54}{2664}{1996}.

\end{thebibliography}
\end{document}